\documentclass[9pt,twocolumn,twoside]{article}

\title{Role of Geometric Shape in Chiral Optics}

\usepackage{amsfonts}
\usepackage{amsmath}
\usepackage{color}
\usepackage{bbold}
\usepackage{hhline}
\usepackage[export]{adjustbox}
\usepackage{bbm}
\usepackage{makecell}
\usepackage{empheq}
\usepackage{cancel}
\usepackage{hyperref}
\usepackage{fp}

\usepackage{verbatimbox}

\newcommand{\epsi}{\varepsilon}
\newcommand{\mue}{\mu}

\newcommand{\Eth}{\vec{\mathcal{E}}}

\newcommand{\Hth}{\vec{\mathcal{H}}}

\newcommand{\Wsca}{W_\text{sca}}

\newcommand{\Ethinc}{\Eth_\text{inc}}
\newcommand{\Hthinc}{\Hth_\text{inc}}

\newcommand{\Ethsca}{\Eth_\text{sca}}
\newcommand{\Hthsca}{\Hth_\text{sca}}

\renewcommand{\vec}[1]{\boldsymbol{#1}}

\newcommand{\colvecTwo}[2]{
    \left( \begin{array}{c}
        #1 \\ #2
    \end{array} \right)}

\newcommand{\xx}{\vec{x}}

\newcommand{\normD}[1]{\left|\left|#1\right|\right|}

\usepackage{pgfplotstable}
\usepackage{pgfplots}
\usepackage{tikz}
\usetikzlibrary{positioning,arrows}
\usetikzlibrary{calc,shapes}

\newcommand{\chiSV}{\chi_\text{SV}}
\newcommand{\chiPW}{\chi_\text{CD}}
\newcommand{\chiTT}{\chi_\text{TT}}
\newcommand{\chiGE}{\chi_\text{GE}}
\renewcommand{\exp}[1]{e^{#1}}

\usepackage{authblk}
\usepackage[a4paper, margin=20mm]{geometry}

\begin{document}

\author[1,2,*]{Philipp Gutsche}
\author[2,3,4]{Xavier Garcia-Santiago}
\author[4]{Philipp-Immanuel Schneider}
\author[5]{Kevin McPeak}
\author[6]{Manuel Nieto-Vesperinas}
\author[2,4]{Sven Burger}
\affil[1]{Freie Universit\"at Berlin, Mathematics Institute, 14195 Berlin, Germany}
\affil[2]{Zuse Institute Berlin, Computational Nano Optics, 14195 Berlin, Germany}
\affil[3]{Karlsruhe Institute of Technology, Institute of Nanotechnology, 76021 Karlsruhe, Germany}
\affil[4]{JCMwave GmbH, 14050 Berlin, Germany}
\affil[5]{Lousiana State University, Department of Chemical Engineering, Baton Rouge, LA}
\affil[6]{Instituto de Ciencia de Materiales de Madrid, Consejo Superior de Investigaciones Cient\'ificas, Madrid, 28049, Spain}

\affil[*]{Corresponding author: gutsche@zib.de}

\twocolumn[
	\begin{@twocolumnfalse}
\maketitle
\begin{abstract}
	The distinction of chiral and mirror symmetric objects is straightforward from a geometrical point of view.
	Since the biological as well as the optical activity of molecules strongly depend on their handedness, chirality has recently attracted high
	interest in the field of nano-optics. Various aspects of associated phenomena including the influences of internal and external
	degrees of freedom on the optical response have been discussed. Here, we propose a constructive method to evaluate the possibility
	of observing any chiral response from an optical scatterer. Based on solely the $T$-matrix of one enantiomer, planes of minimal chiral
	response are located and compared to geometric mirror planes. This
	provides insights into the relation of geometric and optical properties and enables identifying the potential of
	chiral scatterers for nano-optical experiments.
\end{abstract}
  \end{@twocolumnfalse}
]

It is usually a simple task to tell by eye whether an object is chiral or not: Achiral objects are superimposable onto their mirror image
and, accordingly, they possess a mirror plane \cite{kelvin1904}. Recently, chiral scatterers have gained high interest in nano-optics due to
their potential to enhance the weak optical signal of chiral molecules \cite{tang2010,nieto2015,mcpeak2015}. The most common
experimental technique is the analysis of the circular dichroism (CD) spectrum which equals the differential energy extinction due to the illumination
by right- and left-handed circularly polarized light \cite{bohren1940}.

In order to observe such chiral electromagnetic response, it seems to be obvious that geometrically chiral scatterers are required.
However, it has been shown that extrinsic chirality, i.~e. a chiral configuration of the illumination and geometric parameters, yields
comparable effects as intrinsically chiral objects \cite{plum2008}. By tuning the far-field polarization of the illumination, large chiral
near-fields may even be generated in the viscinity of achiral objects \cite{kramer2017}. In CD measurements, randomly orientied molecules are investigated
which can be classified by their $T$-matrix \cite{gutsche2018b}. The latter has been used for quantifying the electromagnetic (e.m.) chirality,
based on a novel definition of it \cite{fernandez2016}.

However, the quantification of the geometric chirality is an elusive task \cite{fowler2005} and even the unambiguous association of the terms right-
and left-handed enantiomer of a general object is impossbile \cite{efrati2014}. Different coefficients attempting to rate the chirality of an object are
based on the maximal overlap of two mirror images \cite{gilat1994} as well as the Hausdorff distance \cite{buda1992}. The choice of a
specific coefficient determines the most chiral object \cite{rassat2004}, i.e.~there is no natural choice for quantifying geometric chirality.
This also holds for the various figures of merit estimating the e.m.~chirality.

In this study, we start by transferring the simple procedure of finding a mirror plane to optics. Namely, we analyse the $T$-matrix and its associated
geometric mirror symmetries by employing translation and rotation theorems of vector spherial harmonics. We illustrate this concept with numerical
simulations of an experimentally realized gold helix. Different quantifications of the e.m.~chirality are compared. Furthermore,
the symmetry planes found in the optical response by our method are correlated to those of geometric origin.
It is shown that the complex optical response, including higher order multipoles, yields mirror planes in the $T$-matrix
which are not directly related to geometric symmetries.

The most general description of an isolated optical scatterer is the well-known $T$-matrix \cite{mishchenko2002}. It relates an arbitrary incident field with the scattered
field caused by the scattering object.
The optical response to any incident field is included in the $T$-matrix.
Accordingly, the following analysis of $T$ is independent of specific illumination parameters such as the direction, polarization
and beam shape. In contrast, the goal of this study is to obtain insights into illumination-independent symmetries of the scatterer.

Usually, both the incident as well as the scattered field are given in the basis of vector spherical harmonics
for computations with the $T$-matrix \cite{jackson1998,appendix}.
Physically observable quantities such as the scattered energy, the absorption, as well as the flux of optical chirality are readily computed from $T$ \cite{gutsche2018b}.
In numerical simulations, $T$ may be computed with high accuracy \cite{garcia2018}. Knowing the response of the left-handed object $T_{l}$ enables the analytic
computation of the response of its mirror image $T_{r}$:
\begin{align}
	T_{r} &= \mathcal{M}_{xy}^{-1} T_{l} \mathcal{M}_{xy},
		\label{eq:Tl}
\end{align}
where we choose mirroring in the $xy$-plane $\mathcal{M}_{xy}$ without loss of generality \cite{appendix}.
Note that the terminology of right- $T_r$ and left-handed $T_l$ is ambiguous, as pointed
out before, and may be interchanged.

\begin{figure}[htbp]
\centering
\includegraphics{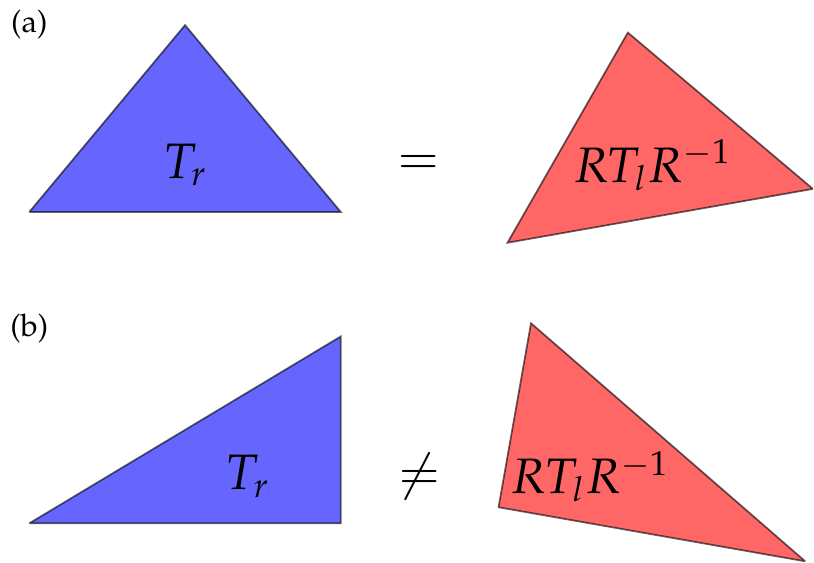}
\caption{
	(a) The mirror image of an achiral object overlaps with its original after proper translations and rotations.
	This implies that the original $T$-matrix $T_{r}$ coincides with $T_{l}$ of the mirror object after the corresponding transformations $R$.
	(b) A chiral object and its mirror image are not congruent. If the object is much smaller than the incident
	wavelength, it usually exists a transformation $R$ after which $T_{l}$ and $R T_{r} R^{-1}$ are equal.
	Note that the achiral isosceles triangle in (a) possess a mirror plane in 2D and that the asymmetric triangle in (b) is chiral only in 2D.
}
\label{fig:chirality}
\end{figure}

Since we aim at investigating arbitrary mirror planes, we note that an arbitrary plane is given by the three spherical coordinates of its normal: the inclination $\Theta$ and the azimuthal angle $\Phi$,
as well as the distance $d$ from the origin.
We define the according transformation $R(\Theta,\Phi,d)$ acting on the object as
\begin{align}
	R(\Theta,\Phi,d) &= \mathcal{T}(\Theta,\Phi,d) \mathcal{R}_{z}(\Phi) \mathcal{R}_{y}(\Theta),
		\label{eq:R}
\end{align}
where $\mathcal{T}(\Theta,\Phi,d)$ is the translation of the $T$-matrix in the direction given by the angles and the distance
and $\mathcal{R}_{z}(\Phi)$ and $\mathcal{R}_{y}(\Theta)$ are the rotations around the $z$- and $y$-axis, respectively \cite{stein1961, appendix}.

For a geometrically achiral object [Fig.~\ref{fig:chirality}(a)], there exists at least one transformation $R(\Theta,\Phi,d)$ such that
$T_{l} = R(\Theta,\Phi,d) T_{r} R^{-1}(\Theta,\Phi,d)$. On the other hand, the lack of a geometric mirror plane of a chiral object [Fig.~\ref{fig:chirality}(b)] implies
that there exists no such transformation and that $T_{l}$ and $T_{r}$ do not coincide for any set of transformation parameters $(\Theta,\Phi,d)$.
Note that this does not generally hold in the long wavelength limit, i.e.~the incident wavelength being much larger than the dimension of the scatterer.

For investigating the role of the geometric shape in nano-optics, it is of interest to identify those planes of highest symmetry of a chiral object:
Although there is no mirror plane in a chiral object, a transformation may be identified in which the right- and left-handed $T$-matrices are closest to one another.
Rating the closeness is done here by calculating the 2-norm of the difference of these two matrices. Accordingly, we introduce the coefficient $\chiTT$ which
minimizes the difference between the $T$-matrices of mirror images as
\begin{equation}
	\chiTT = \min_{(\Theta, \Phi, d)} \normD{ T_{l} - R^{-1}(\Theta, \Phi, d) T_{r} R(\Theta, \Phi, d) }_{2}.
	\label{eq:min}
\end{equation}
This means that for the mirror plane corresponding to minimal parameters
$(\Theta_{\text{min}},\Phi_{\text{min}},d_{\text{min}})$
of \eqref{eq:min}, the optical responses of the two mirror images are as similar as possible. In other words, the mirror images are hardly distinguishable.
For an achiral object $\chiTT$ vanishes since there exists a transformation for which the mirror images are identical.

Obviously, the choice of the norm is not unique and other quantifications of similarity of the mirror images could be defined (cf.~\cite{appendix} for the physical relevance of the 2-norm).
A recently introduced coefficient $\chiSV$ is e.g.~based on the singular-value decomposition of the $T$-matrix in the helicity basis \cite{fernandez2016}.
Alternatively, the angular-averaged differential energy extinction $\chiPW$ due to illuminating with either right- or left-handed circularly polarized plane waves
is experimentally accessible as the CD spectrum.

\begin{figure}[htbp]
\centering
\includegraphics{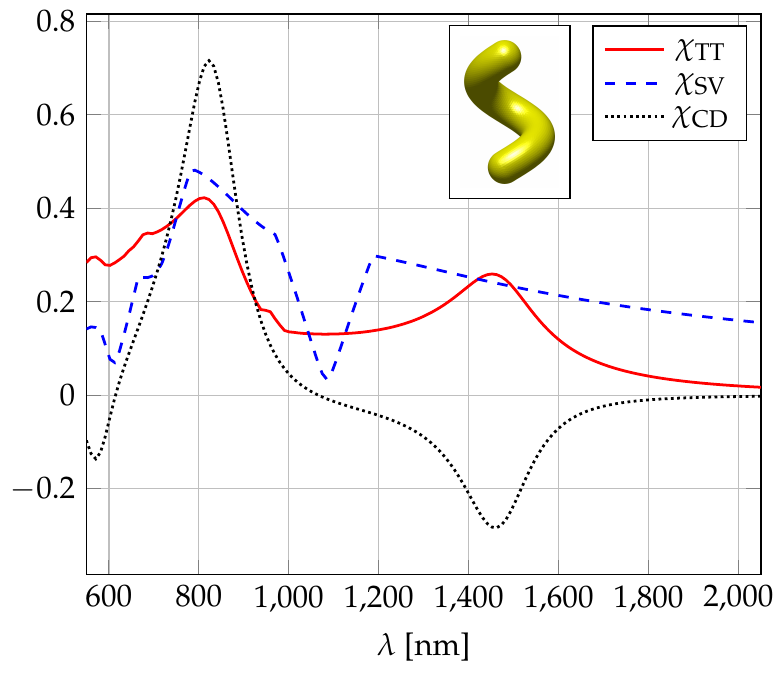}
\caption{
	Chiral response of a gold nano-helix depending on the incident wavelength $\lambda$.
	The angular averaged differential extinction of circularly polarized plane waves $\chiPW$ (black dotted line) vanishes at $615\text{nm}$ and $1070\text{nm}$
	which could be interpreted as achirality of the studied object.
	The electromagnetic chirality coefficient $\chiSV$ (dashed blue line) is based on the singular values of the $T$-matrix
	in the helicity basis. Values below $0.1$ at $610\text{nm}$ and $1085\text{nm}$ indicate nearly achiral optical response.
	However, the minimal difference $\chiTT$ (red solid line) between $T_{r}$ and $R T_{l} R^{-1}$ reveals that the helix is chiral at all wavelengths. Its maxima
	correspond to those of $\chiPW$ and are, hence, observable.
}
\label{fig:spectrum}
\end{figure}

In Figure \ref{fig:spectrum}, we compare simulations of these three coefficients quantifying the e.m.~chirality for a gold helix as realized experimentally \cite{wozniak2018}.
The helix is constructed on the surface of a cylinder \cite{appendix} with height 230nm and radius 60nm.
The CD spectrum $\chiPW$ shows zero values at incident wavelengths of $\lambda=615\text{nm}$ and $\lambda=1070\text{nm}$. If only these wavelengths were analyzed, one could draw
the conclusion that an achiral object is investigated. Nevertheless, CD makes the chiral geometric nature of the helix visible as a maximum at $823\text{nm}$ and
a minimum at $1,452\text{nm}$ of smaller amplitude. For a helix with an opposite twist, i.e~the mirror image, the roles of the extrema are interchanged.

On the other hand, the coefficient $\chiSV$ is normalized by the average interaction strength of the $T$-matrix at each wavelength. This yields a fairly flat spectrum with two narrow
minima below $0.1$ at the two $\lambda$ for which $\chiPW=0$.
These minima are not present in the minimized $\chiTT$ introduced in \eqref{eq:min}. However, the maxima of this latter coefficient are in accordance with the experimentally observable
CD extrema ($\chiPW$). In the long wavelength regime, all three coefficients tend to zero as expected for point-like particles due to vanishing off-diagonal elements in the $T$-matrix.

\begin{figure}[htbp]
\centering
\includegraphics{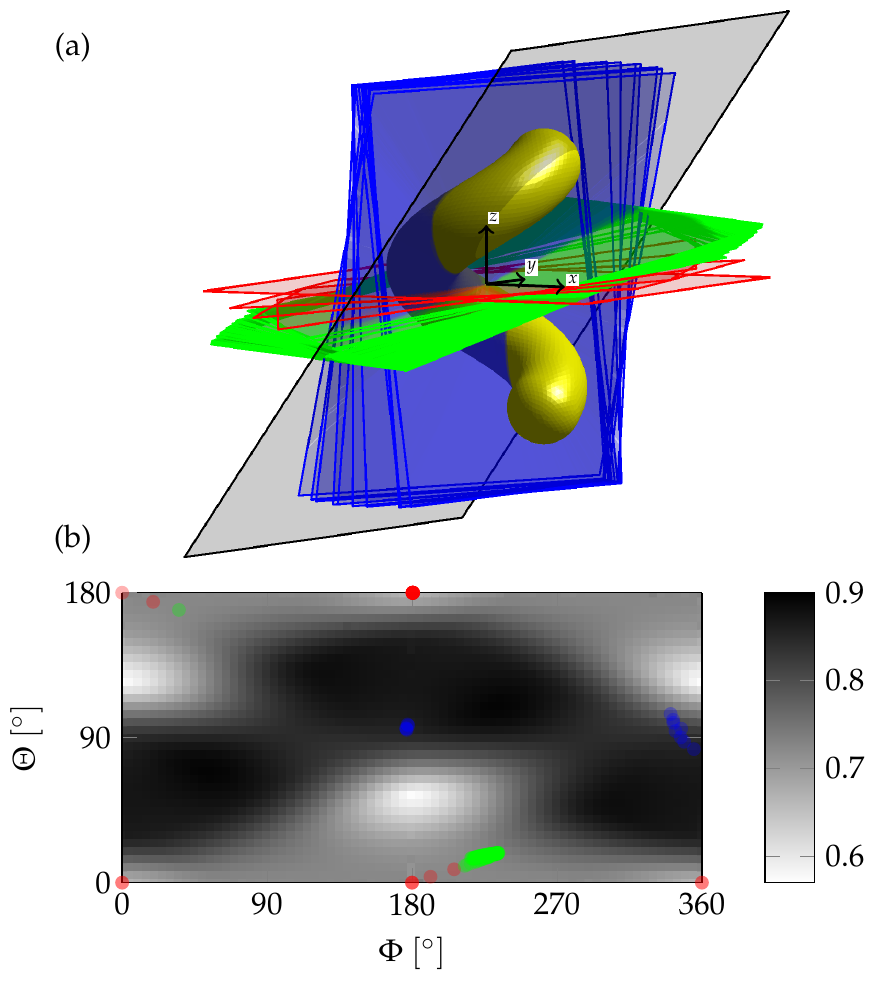}
\caption{
	(a) Transformed $xy$-planes (blue, red, green) corresponding to minimal $\chiTT$ computed from $T$-matrix of the gold helix (yellow).
	The dark grey plane corresponds to minimal $\chiGE$.
	(b) Geometric chiral coefficient $\chiGE(\Theta,\Phi)$ for the helix and its mirror image which is rotated around the centroid (grey colormap).
	The minimal value of 0.57 belongs to the dark grey plane in Fig.~\ref{fig:planes}(a). Angles of the colored planes are shown by circles.
}
\label{fig:planes}
\end{figure}

The minimization in the three-dimensional parameter space in \eqref{eq:min} is carried out using Bayesian optimization \cite{schneider2017, appendix}. Since the shape
of the minimized function highly depends on the actual object, the Bayesian approach is well suited for finding a global minimum. The parameters
$(\Theta_{\text{min}},\Phi_{\text{min}},d_{\text{min}})$
of the optimized value are related to geometric mirror planes.
In Figure \ref{fig:planes}(a), the planes following from the respective transformation
$R(\Theta_{\text{min}},\Phi_{\text{min}},d_{\text{min}})$ of the $xy$-plane are plotted for all incident wavelengths from $550\text{nm}$ to $2.05\mu\text{m}$.
The inclination $\Theta$ and azimuthal angle $\Phi$ are given in the shown coordinate system which is centered at the centroid of the helix.

We identify three distinct classes shown in blue, red and green. These correspond to planes which are parallel and perpendicular to the helix axis, as well as
tilted by a small angle $\Theta$ from the horizontal position, respectively.
The dark grey plane corresponds to the minimal geometric parameters which will be explained in the following paragraphs.
Details on the optimization such as challenging flat behaviour for translations from the centroid, and on
the obtained minimizing parameters, are given in \cite{appendix}.

Next, we compare the findings on the symmetry based on the optical $T$-matrix to those stemming from purely geometric properties.
As discussed previously, there is no coefficient which unambiguously rates the geometric chirality of an object.
We choose a coefficient $\chiGE$ based on the overlap of the left- $O_{l}$ and right-handed $O_{r}(\Theta,\Phi,d)$ object,
where the latter results from mirroring $O_{l}$ at the $xy$-plane and transformation with $(\Theta,\Phi,d)$.
Namely, the volume $V$ of the overlap is compared to the volume of the object \cite[Eq.~(8) in]{gilat1994,appendix}:
\begin{equation}
	\chiGE(\Theta,\Phi,d) = 1 - \frac{V\left(O_{l} \cap O_{r}(\Theta,\Phi,d)\right)}{V(O_{l})}.
	\label{eq:chige}
\end{equation}
This coefficient vanishes for achiral objects as required for a degree of chirality \cite{buda1992}.

Figure \ref{fig:planes}(b) displays the geometric chirality coefficient $\chiGE(\Theta,\Phi,0)$ for planes rotated around the centroid of the helix as a grey colormap.
Dark regions with large values of $\chiGE$ indicate a vanishing overlap between the two mirrored helices. Note that for large distances to the origin $d\rightarrow\infty$,
the mirror images do not overlap and $\chiGE=1$. However, this is always possible no matter if the object is chiral or not. As in the case of $\chiTT$,
the parameter points of interest of $\chiGE(\Theta,\Phi,d)$ are those corresponding to a minimum: The minimum $0.57$ in Fig.~\ref{fig:planes}(b) occurs at $(180^\circ,55^\circ)$ and 
$(0^\circ,125^\circ)$ which show the instrinic chiral property of the investigated helix. These two minima are equivalent since a finite helix is $C_2$ symmetric.
The corrsponding transformed $xy$-plane is shown in dark grey in Fig.~\ref{fig:planes}(a).

Alongside the geometric coefficient $\chiGE$, the planes identified for the minimized $T$-matrix difference are shown as colored circles in Fig.~\ref{fig:planes}(b).
The colors (red, blue and green) of these circles are the same colors used for the planes, i.e.~a direct comparison of the angle parameters is possible.
As seen, the planes are ranked according to their $\Theta$ values: The perpendicular class 1 (blue) has $\Theta \in [83, 105]^\circ$
The flat planes belonging to class 2 (red) show $\Theta \in [0,8.5]^\circ$ and $\Theta \in [174, 180]^\circ$ and the tilted class 3 (green) has $\Theta \in [10, 19]^\circ$ and $\Theta = 170^\circ$.

\begin{figure}[htbp]
\centering
\includegraphics{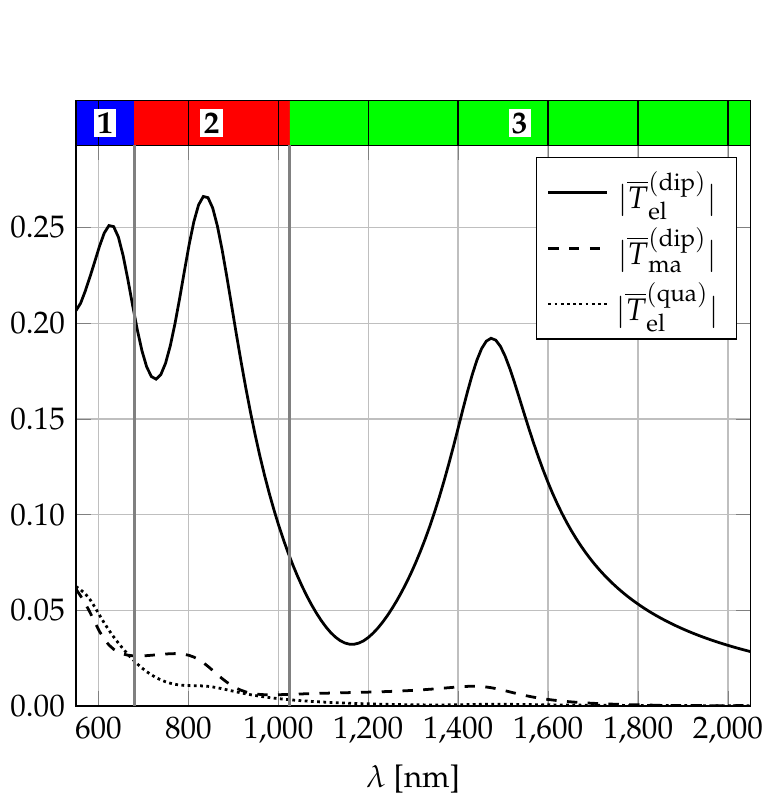}
\caption{
	Wavelength-dependent classification of symmetry planes of $T$-matrix (top).
	Absolute value of averaged diagonal $T$-matrix entries corresponding to induced electric dipoles (solid), magnetic dipoles (dashed) and electric quadrupoles (dotted).
	The classes 3 (green), 2 (red) and 1 (blue) belong to decreasing wavelengths. Changes in symmetry of $T$ are due to higher order multipoles.
}
\label{fig:multclass}
\end{figure}

None of the three optical symmetry planes is directly related to the geometric mirror plane of the helix.
However, Fig.~\ref{fig:planes}(b) enables the comparison of geometric and optical symmetries.
In order to further analyze the optical response, we show the wavelength-dependent classification of the symmetry planes on top of Fig.~\ref{fig:multclass}.
The three classes correspond to sharply separated wavelength ranges: Class 1 is valid for $\lambda \in [550,680]\text{nm}$. For larger wavelengths $\lambda \in [680,1025]\text{nm}$,
the $T$-matrix possesses the symmetry according to planes of class 2. Finally, in the long wavelength regime ($\lambda \in [1025, 2050]\text{nm}$), the symmetry is in class 3.

The analysis in Fig.~\ref{fig:planes}(b) suggests that class 3 (green) is the closest one to the geometric mirror plane. This is further strengthened by the full angular spectrum
of the optical chirality coefficient $\chiTT$ \cite{appendix}. Accordingly, we find that the optical response is dominated by the geometric shape for long wavelengths.
Obviously, the optics is dominated by the electric dipole moment in this regime which is also shown in Fig.~\ref{fig:multclass}. Here, the mean of the diagonal entries of submatrices of
the $T$-matrix are shown. These are proportional to the electric and magnetic dipole moments as well as to the electric quadrupole moments.

The three symmetry classes of the $T$-matrix occur close to three electric dipole peaks
($\lambda=623,~833,~\text{and}~1,473\text{nm}$) and are influenced by the anisotropy of the $T$-matrix. Truly chiral behaviour as observed here, however, originates not from anisotropy but
from coupling between electric and magnetic multipoles \cite{zambrana2016}. In the appendix we elaborate on the complex interplay between these different contributions in the dipolar
limit \cite{appendix}. Here, we limit the discussion to the main aspects of different multipolar contributions.

For large wavelengths with symmetry of class 3, the electric dipoles are much larger then any other induced multipole.
In the intermediated regime of symmetry class 2, the magnetic dipole moment significantly increases. For short wavelengths with planes of class 1, the electric quadrupole moment
is stronger than the magnetic dipole moment which yields the change in the optical symmetry.
Higher order multipoles including mixed electric-magnetic moments are depicted in \cite{appendix}.
The more elaborate study of multipolar resonances underlines again that the chiral response deviates from expectations due to a purely geometrical analysis of the scatterer.

In summary, we have introduced a method to obtain geometric mirror planes from the optical $T$-matrix of a scattering object. We applied this method to an isolated gold helix
and found correlations between the symmetry of its geometric shape and those of the optical response in the long wavelength regime.
On the one hand, this confirms the expectation that instrinsic geometric chirality is directly related to an optically chiral response. On the other hand, for shorter wavelengths where higher
multipoles are induced, mirror planes derived from the $T$-matrix do not coincide with the geometric mirror plane.
This implies light-matter interactions whose symmetry cannot be explained simply by geometric chirality.
Our method can be applied to all isolated scattering objects and constructively identifies planes of mirror symmetry.



\textbf{Funding} We acknowledge support from Freie Universit\"at Berlin through the Dahlem Research School.
This project has received funding from the European Union’s Horizon 2020 research and innovation programme under the Marie Sklodowska-Curie grant agreement No 675745.
M.~N.-V. is supported by MINECO grants FIS2014-55563-REDC, and FIS2015-69295-C3-1-P.

\textbf{Acknowledgements} We thank Martin Hammerschmidt for in-depth discussions on several topics.



\appendix

\section{Appendix}

\subsection{Vector Spherical Harmonics}
The solution of Maxwell's equations for an isolated scatterer being the subject to external illumination is conventiently
expressed in the basis of vector spherical harmonics (VSHs) $\vec{N}_{nm}^{(l)}(\xx)$ and $\vec{M}_{nm}^{(l)}(\xx)$ \cite{mishchenko2002}.
The index $n$ is the multipole order and gives the total angular momentum.
The index $m$ is related to the eigenvalue $m (m+1)$ of the squared orbital angular
momentum operator $L^2$ \cite[Sec.~9.7,]{jackson1998}. The superscript $l$ differentiates between incident ($l=1$) and
scattered ($l=3$) electromagnetic fields.

In the VSH basis, the incident electric $\Ethinc$ and magnetic $\Hthinc$ time-harmonic fields are
\begin{align}
	\Ethinc(\xx,t) &= \exp{-i \omega t} \sum_{n=1}^{\infty} \sum_{m=-n}^{m=n} p_{mn} \vec{N}_{mn}^{(1)}(\xx) + q_{mn} \vec{M}_{mn}^{(1)}(\xx)
	, \\
	\Hthinc(\xx,t) &= -\frac{i \exp{-i \omega t}}{Z} \sum_{n=1}^{\infty} \sum_{m=-n}^{m=n} p_{mn} \vec{M}_{mn}^{(1)}(\xx) + q_{mn} \vec{N}_{mn}^{(1)}(\xx)
	,
\end{align}
with the wave impedance $Z=\sqrt{\mue_0\mue/(\epsi_0\epsi)}$ and the relative permeability $\mue$ and relative permittivity $\epsi$
of the surrounding medium.
The scattered fields obeying the radiation condition are given by
\begin{align}
	\Ethsca(\xx,t) &= \exp{-i \omega t} \sum_{n=1}^{\infty} \sum_{m=-n}^{m=n} a_{mn} \vec{N}_{mn}^{(3)}(\xx) + b_{mn} \vec{M}_{mn}^{(3)}(\xx)
	, \\
	\Hthsca(\xx,t) &= -\frac{i \exp{-i \omega t}}{Z} \sum_{n=1}^{\infty} \sum_{m=-n}^{m=n} a_{mn} \vec{M}_{mn}^{(3)}(\xx) + b_{mn} \vec{N}_{mn}^{(3)}(\xx)
	.
\end{align}
The VSHs $\vec{N}_{nm}^{(3)}(\xx)$ and $\vec{M}_{nm}^{(3)}(\xx)$ are the electric fields due to induced electric and magnetic multipoles, respectively.

\subsection{$T$-Matrix}
The coefficients $p_{mn}$ and $q_{mn}$ as well as $a_{mn}$ and $b_{mn}$ are the VSH coefficients of the incident and scattered field, respectively.
The optical response of a scatterer is described by the relation of these two sets of coefficients. Specifically, the $T$-matrix enables the
computation of the scattered field $(\vec{a},\vec{b})$ from a known incident field $(\vec{p}, \vec{q})$, where all coefficients are summarized in vectors:
\begin{align}
	T \colvecTwo{\vec{p}}{\vec{q}} = \colvecTwo{\vec{a}}{\vec{b}}. 
\end{align} 
Accordingly, all possible observable quantities such as scattered energy, absorption and chirality extinction can be deduced
from the $T$-matrix \cite{mishchenko2002, gutsche2018b}.

\begin{figure}[htbp]
\centering
\includegraphics{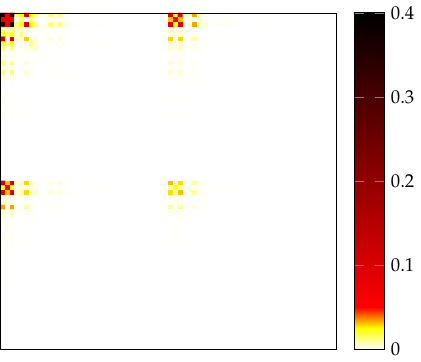}
\includegraphics{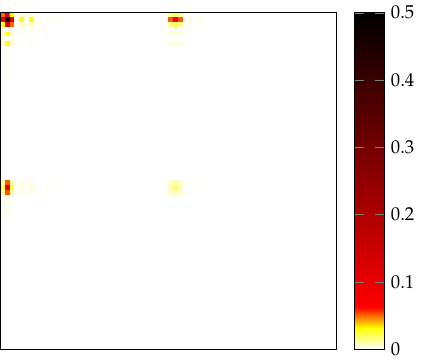}
\caption{
	Colorplot of absolute values of the $T$-matrix entries at $\lambda=823\text{nm}$ (left) and $\lambda=1,452\text{nm}$ (right) which correspond to extremal values in the CD spectrum.
}
\label{fig:T}
\end{figure}

In this study, we analyze quadratic $T$-matrices of dimension $2 N (N+2)$ with maximal multipole order $N=5$.
$T$ is obtained numerically by the illumination with 150 plane waves. The respective incident wave vectors are distributed equidistanly on a spherical surface
and the polarizations are chosen randomly. The projection of the scattered field onto VSHs is computed from general surface integrals \cite{garcia2018}.
The resulting matrices for the extremal CD response (Figure \ref{fig:spectrum}) at $\lambda=823\text{nm}$ and $\lambda=1,452\text{nm}$ are shown in Figure \ref{fig:T}.
 
\subsection{Transformation of $T$}
The $T$-matrix is useful for transformations $R$ such as rotating $\mathcal{R}$, translating $\mathcal{T}$ and mirroring $\mathcal{M}$
the object of interest. Under mirror reflection $\mathcal{M}_{xy}$ on the $xy$-plane, both the incident ($l=1$) and the scattered ($l=3$) VSHs transform as
\begin{align}
	(\mathcal{M}_{xy} \vec{N}_{mn}^{(l)}) (\xx) &= (-1)^{(m+n)} \vec{N}_{mn}^{(l)}(\xx)
	,\\
	(\mathcal{M}_{xy} \vec{M}_{mn}^{(l)}) (\xx) &= (-1)^{(m+n+1)} \vec{M}_{mn}^{(l)}(\xx)
	.
\end{align}
Accordingly, the matrix $T_l$ representing the mirror image of the scatterer with $T$-matrix $T_r$ is given by
\begin{align}
	T_l = \mathcal{M}_{xy}^{-1} T_r \mathcal{M}_{xy},
\end{align}
with $(\mathcal{M}_{xy})_{ij} = (-1)^{(m+n)} \delta_{ij}$ for $i=1,...,N(N+2)$ and
with $(\mathcal{M}_{xy})_{ij} = (-1)^{(m+n+1)} \delta_{ij}$ for $i=N(N+2)+1,...,2N(N+2)$.

\begin{figure}[htbp]
\centering
\includegraphics{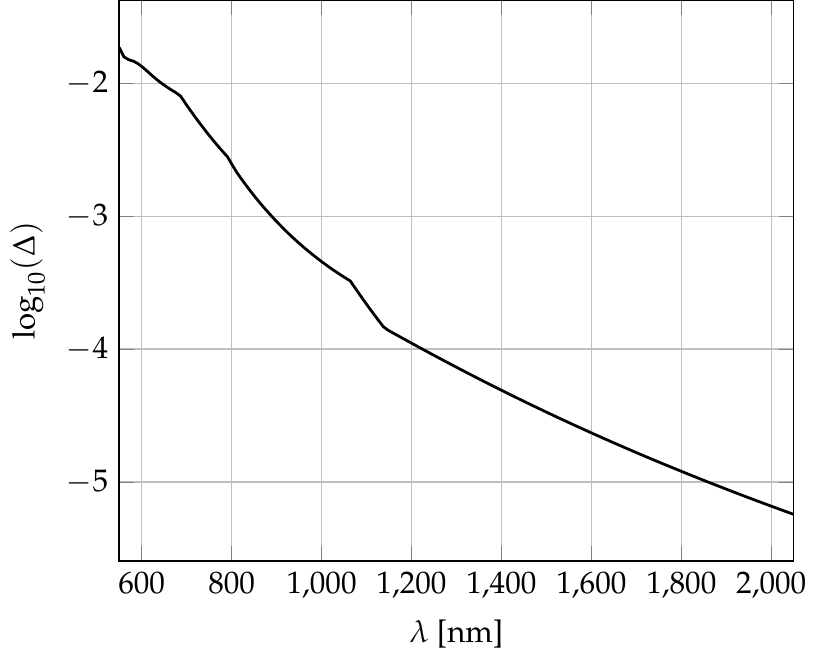}
\caption{
	Numerical error of translation due to the finite size of $T$ with maximal multipole order $N=5$.
}
\label{fig:transl}
\end{figure}

By employing the addition theorems for translation $\mathcal{T}$ and rotation $\mathcal{R}$ of VSHs \cite{stein1961}, the $T$-matrix of the transformed
scatterer is computed analytically. Note that the mirrored and rotated $T$-matrices are exact, whereas the translated $T$-matrix
is an approximation due to the finite size of $T$, i.e.~the maximal multipole order $N=5$.
In Figure \ref{fig:transl}, we plot the error $\Delta$ of forward and backward translation in the same direction with $\Theta=76^\circ, \Phi=330^\circ$ and
$d=206\text{nm}$. Although the scatterer is mapped onto its original position, the truncation of $T$ introduces the error
\begin{align}
	\Delta = \frac{
		\max_{ij} \left|
		\left( T_r -
			\mathcal{T}^{-1}_{\text{back}} \left\{ \mathcal{T}^{-1}_{\text{forw}}
			T_r
			\mathcal{T}_{\text{forw}} \right\} \mathcal{T}_{\text{back}}
		\right)_{ij}
		\right|
		}{
		\max_{ij} \left| \left( T_r \right)_{ij} \right|
		}.
\end{align}

\subsection{Mirror Planes in $T$}
\begin{figure}[htbp]
\centering
\includegraphics{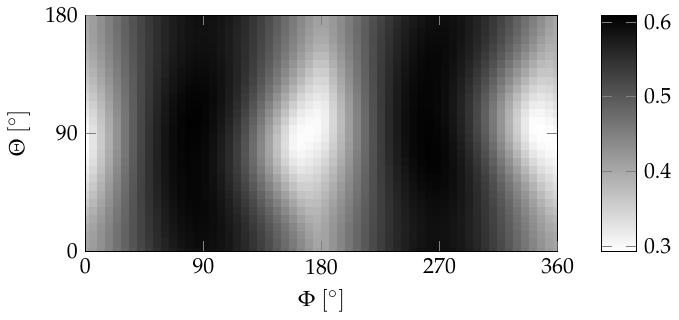}
\includegraphics{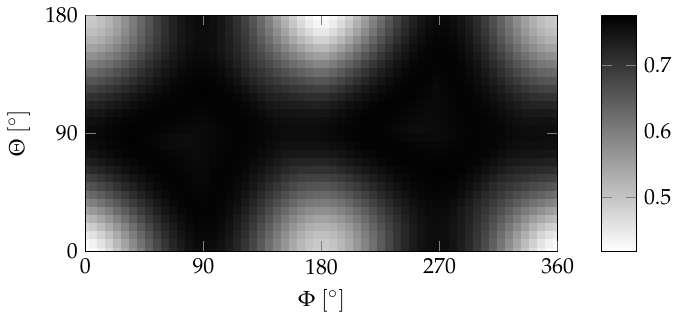}
\includegraphics{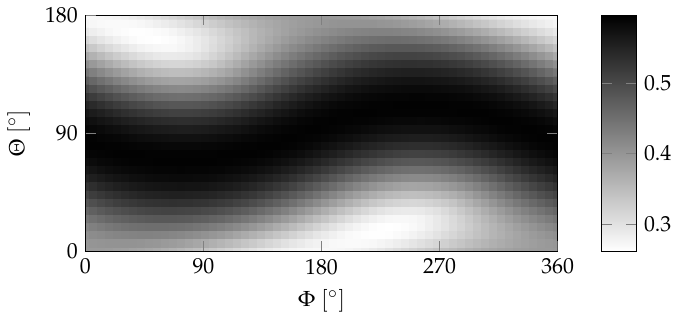}
\caption{
	$\chiTT(\Theta,\Phi,0)$ is the norm of the difference between original $T$-matrix and the matrix of a mirrored scatterer rotated by $\Theta$ and $\Phi$ around the centroid
	of the helix for $\lambda=623\text{nm}$ (top), $\lambda=823\text{nm}$ (middle) and $\lambda=1,452\text{nm}$ (bottom).
}
\label{fig:angleT}
\end{figure}

\begin{figure}[htbp]
\centering
\includegraphics{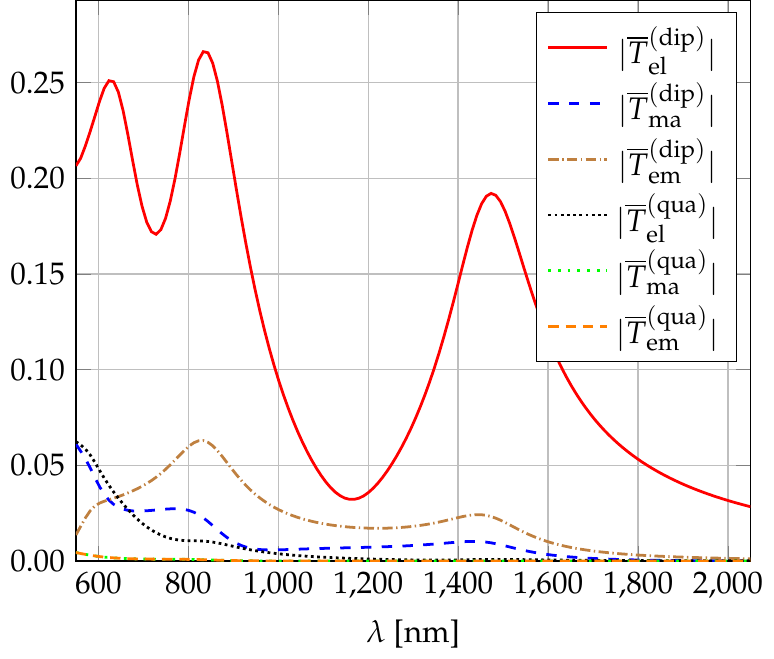}
\caption{
	Absolute value of averaged entries of $T$ corresponding to electric, magnetic and electric-magnetic dipole and quadrupole moments.
}
\label{fig:multipoles}
\end{figure}

As discussed in the main text, the aim of this study is to find symmetries in $T$ which correspond to geometric mirror planes, or at least
identifying planes of highest possible mirror symmetry for which the $T$-matrices of the original scatterer and its mirror image show the highest similarity.
This is done by the global minimization in \eqref{eq:min}.
As illustrative examples, we show the angle-dependent, i.e.~non-minimized, $\chiTT(\Theta,\Phi,0)$ for rotations by angles $\Theta$ and $\Phi$ without translation ($d=0$),
in Figure \ref{fig:angleT}. All possible rotations of two matrices at the extremal values of the CD spectrum are shown (cf.~Figure \ref{fig:T}) with
\begin{align}
	\chiTT(\Theta, \Phi, d) = \normD{ T_{l} - R^{-1}(\Theta, \Phi, d) T_{r} R(\Theta, \Phi, d) }_{2}.
\end{align}
It is apparent in the change of similarity planes (Figure \ref{fig:planes}) and the angular dependence of the similarity
of the $T$-matrices of mirror images (Figure \ref{fig:angleT}) that the symmetry of $T$ is highly wavelength dependent.
This is due to the fact that for shorter wavelengths, higher multipoles contribute to the overall response as shown in Figure \ref{fig:multipoles}.
There, we show the absolute value of the $T$-matrix entries which correspond to averaged electric and magnetic multipole orders $N=1$ (dipole) and $N=2$ (quadrupole).
Overall the response is dominated by electric dipole contributions. Below $\lambda=1\mu\text{m}$, magnetic dipole and electric quadrupole
effects are increasing and result in symmetry planes which are not found in a purely geometric analysis.

\subsection{Dipolar Analysis of Symmetry Classes}

\begin{figure}[htbp]
\centering
\includegraphics{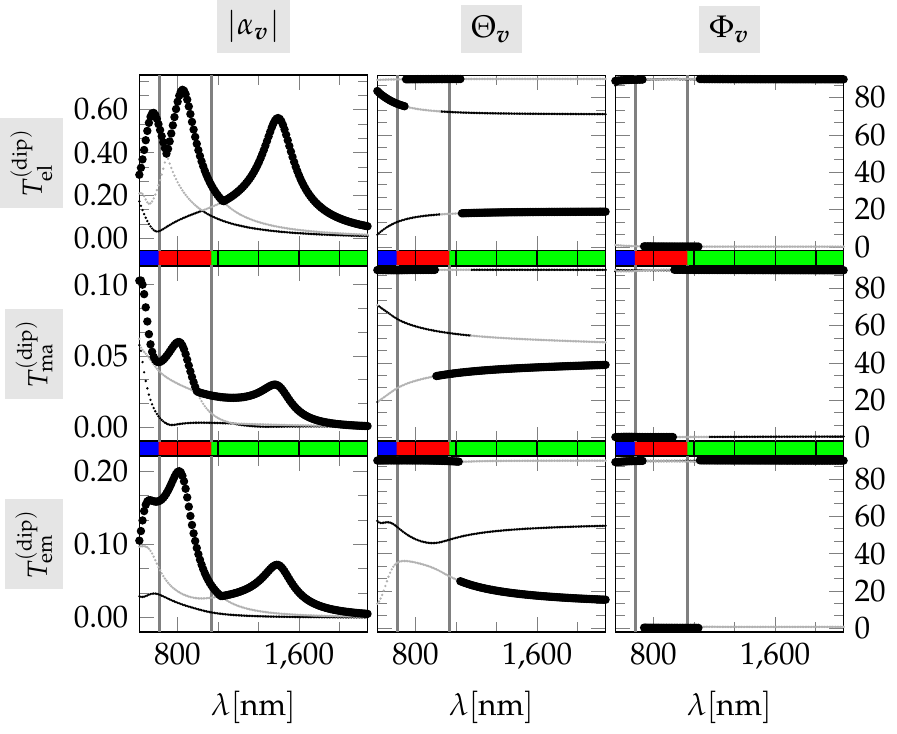}
\caption{
	Eigenvalue analysis of dipolar contributions. See section on dipolar analysis for details.
}
\label{fig:eigen}
\end{figure}

In order to investigate the physical origin of the three symmetry classes found in our analysis, we further study the $T$-matrix in the dipolar approximation, i.e.~
the electric $T_{\text{el}}^{\text{(dip)}}$, the magnetic $T_{\text{ma}}^{\text{(dip)}}$ and the electric-magnetic part $T_{\text{em}}^{(dip)}$.
For each 3x3 matrix, we compute the three eigenvalues $\alpha_{\vec{v}}$ with e.g.~$T_{\text{el}}^{\text{(dip)}} \vec{v} = \alpha_{\vec{v}} \vec{v}$.
In Fig.~\ref{fig:eigen}, we show the absolute value of $\alpha_{\vec{v}}$ as well as the spherical coordinates $\Theta_{\vec{v}}$ and $\Phi_{\vec{v}}$ of the
respective eigenvector.

The quantities belonging to the largest eigenvalue are plotted with large black circles. The second largest eigenvalue is depicted with small black circles and the smallest one
has small gray circles. The $y$-axis of $|\alpha_{\vec{v}}|$ is presented on the left and the $y$-axis for the two angles is placed on the right.
Alongside the $x$-axis, the colored wavelength-dependent symmetry classes found in our study are shown. The transitions between these classes at $\lambda=680\text{nm}$ and $\lambda=1025\text{nm}$
are shown with gray vertical lines.

The transitions between the symmetry classes occur for a change in the dominant eigenvalue of the electric, magnetic and electric-magnetic submatrices, i.e.~for different resonances of the
scatterer. That is why the symmetry planes discussed in the main text are sharply separated. There is always a smooth line connecting the parameters of the eigenvalues,
however, the dominant eigenstate is highly wavelength-dependent and introduces the drastic changes in the symmetry of $T$.

Note that the changes of the maximal eigenvalue do not exactly coincide for the electric, magnetic and electric-magnetic matrices.
This illustrates that the symmetry of $T$ and especially the chiral behaviour of the helix is dependent on a complex interplay between electric and magnetic contributions as well
as higher order multipoles which are not shown here.
In comparison with the multipolar resonance behavior, the scalar coefficient $\chiTT$ introduced in the main text largely simplifies the analysis of the chiral behavior of the scatterer.

\subsection{Physical Relevance of the 2-norm}
The electromagnetic chirality coefficient $\chiPW$ is directly accessible by experiment: it is the differential energy extinction due to the illumination
by right and left handed circularly polarized plane waves. The observed energy extinctions at different angles of incidence $(\Theta,\Phi)$ are averaged
over the whole solid angle. Numerically, this quantity is deduced from the $T$-matrix which contains the full angular- and polarization-resolved optical
response of the scatterer \cite[Eq.~(27)]{gutsche2018b}.

For the coefficient $\chiSV$ the $T$-matrix itself is analyzed rather than the experimentally measurable quantities in the experiment.
Here, the ability of retrieving the full optical response by illumination with fields of only one state of well-defined helicity is quantified \cite{fernandez2016}.
Irrespective of any particular conditions of the incident field and/or the experimental observables, a general property of the scattering object is obtained,
namely, whether the scatterer is electromagnetically chiral. The latter property is introduced in Ref.~\cite{fernandez2016}.

In the current study, we put forward a concept relating the geometric property of (a)chirality of an object to its optical properties. Specifically,
we find planes of similarity in the $T$-matrix which directly relate to geometric mirror planes. For simplicity, we choose the 2-norm $\normD{\cdot}_2$ in \eqref{eq:min}.
Depending on the experimental setup, other matrix norms or operations may be chosen which correspond to observables such as the scattered chirality \cite{gutsche2018a}.
In any case, the measurable quantity can be computed from the $T$-matrix since all optical information is contained therein.

For the specific choice of the 2-norm, not only the geometric mirror plane of highest similarity between mirror images is obtained.
Further, the required illumination parameters are given directly. We recall that the 2-norm of any matrix $A$ is its maximal singular value $\sigma_{\text{max}}$: $\normD{A}_2 = \sigma_{\text{max}}(A)$.
Additionally, the singular value decomposition is a factorization of $A$ into $A=U \Sigma V^*$ with the diagonal matrix of singular values $\Sigma$ and unitarity matrices $U$ and $V$.
The matrices $U$ and $V$ consist of the left- $\vec{u}$ and right-singular vectors $\vec{v}$ of singular value $\sigma$ with
\begin{align}
	A \vec{v} = \sigma \vec{u}.
\end{align}
Note that due to the unitarity of $U$ and $V$, the singular vectors are normalized: $\normD{\vec{u}}_2 = \normD{\vec{v}}_2 = 1$.

In \eqref{eq:min}, the difference of the matrices $T_l$ and $T_r$ of the (transformed) mirror images are compared. Accordingly, it holds $\chiTT = \sigma_{\text{max}}(T_r - R^{-1} T_l R)$,
where we omit the rotation and translation parameters $(\Theta,\Phi,d)$. The singular value decomposition gives the VSHs coefficients of the incident field with minimal $\chiTT$
as $\vec{v}_{\text{max}} = (\vec{p}_{\text{max}}, \vec{q}_{\text{max}})$ with
\begin{align}
	\chiTT = \normD{(T_r - T_l) \colvecTwo{p_{\text{max}}}{q_{\text{max}}}}_2
		= \Wsca(T_r - T_l).
\end{align}
The last step is fulfilled since the absolute value of the scattered VSH coefficients is proportional to the scattered energy \cite{mishchenko2002, gutsche2018a}.
This means the coefficient $\chiTT$ is proportional to the scattered energy of the differential field due to the mirror images $T_r$ and $T_l$ with the smallest discrepancy.
The respective geometric parameters $(\Theta_{\text{min}}, \Phi_{\text{min}}, d_{\text{min}})$ of $\chiTT$ follow from the optimization parameters.
Accordingly, we obtain in our approach both the incident field (from its VSH coefficients) as well as the mirror plane (from its parameters) corresponding to an experimental realizable setup.

\subsection{Geometric Model}
The analyzed object is a gold helix with paramters taken from Ref.~\cite{wozniak2018}. The permittivity of Au is derived from a fit of experimental data
to a Lorentz-Drude model \cite{rakic1998}.

\begin{figure}[htbp]
\centering
\includegraphics[width=.3\linewidth]{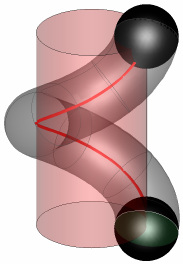}
\caption{
	Construction of helix: Spheres (black) are placed at the top and bottom. A circle (green) is swept along a line (red) on the
	cylindrical surface (red, transparent). The helix (grey, transparent) consists of six parts in parameter space.
}
\label{fig:helixParts}
\end{figure}

In Fig.~\ref{fig:helixParts}, the construction of the helix based on a CAD-model is shown.
Two spheres with radius of 35nm are placed at the top and bottom of a cylinder with radius of 60nm and height of 230nm.
This yields $z=\pm 115\text{nm}$ and $x=60\text{nm}$ for the upper and lower sphere, respectively.
A circle of radius 35nm is swept along a path on the cylindrical surface. Due to numerical stability, the path is divided into six segments.
This procedure results in a helix with one coil and a circular cross section. 

\subsection{Geometric Chirality Coefficient}
\begin{figure}[htbp]
\centering
\includegraphics[width=.3\linewidth]{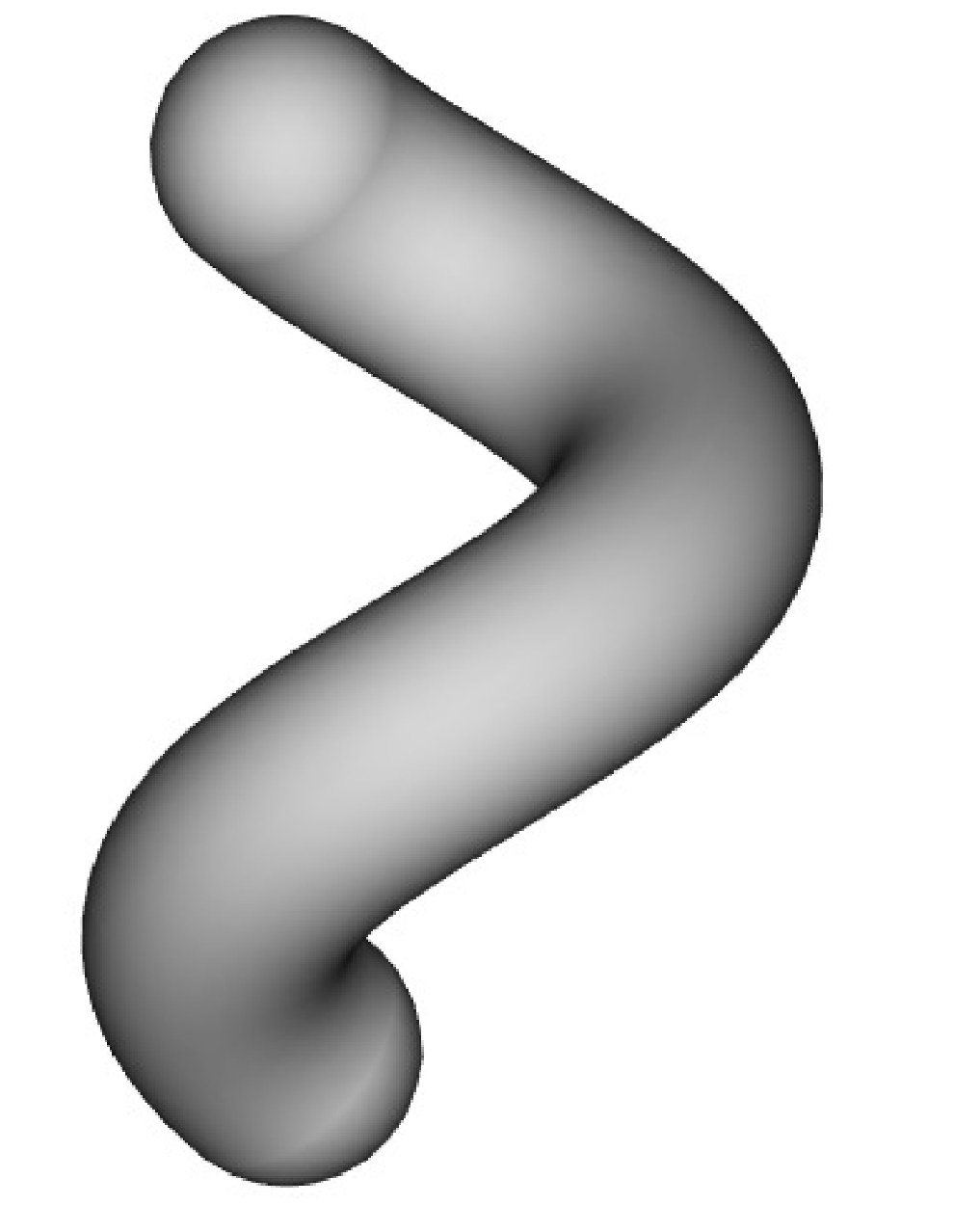}
\includegraphics[width=.3\linewidth]{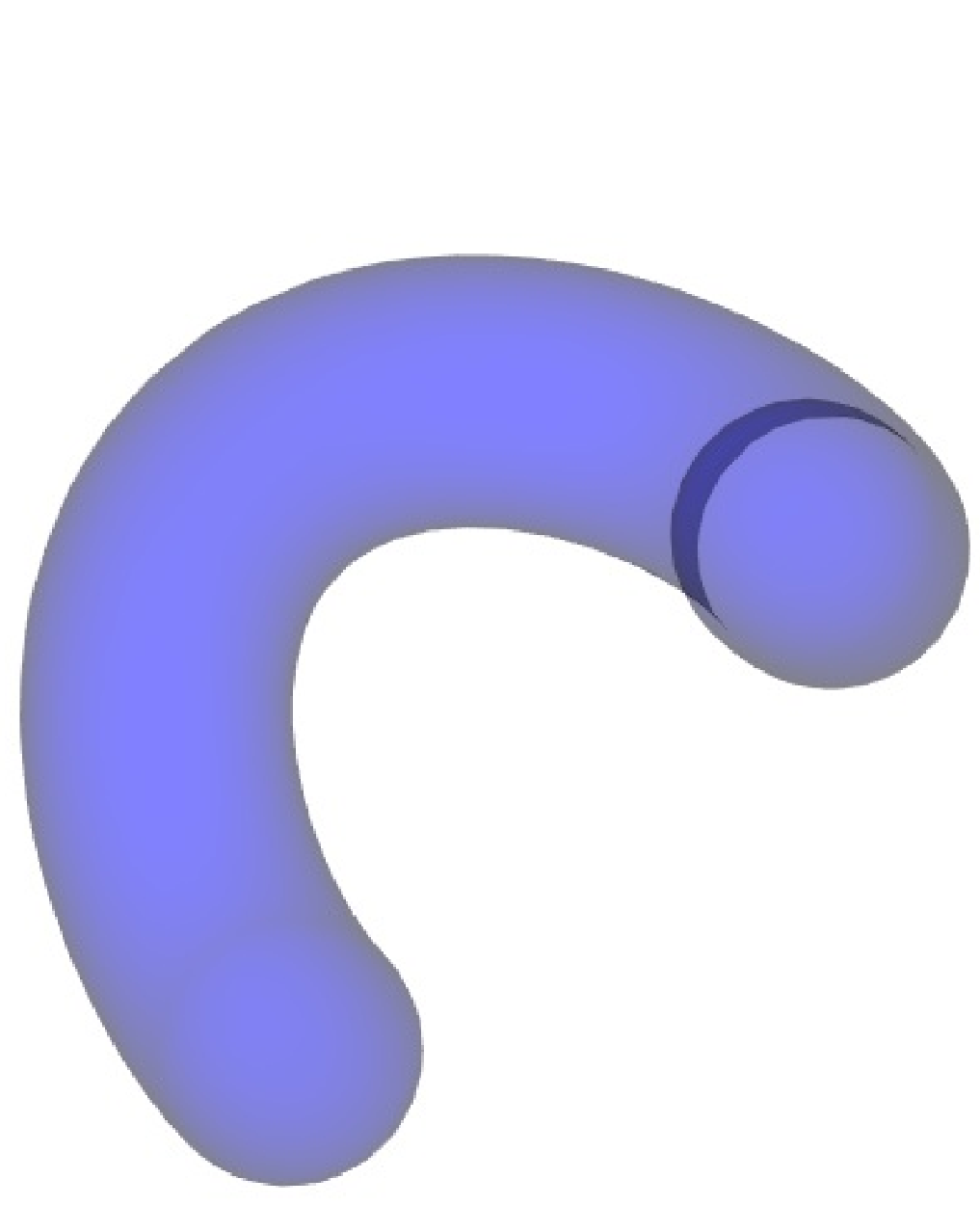}
\\
\vspace{2ex}
\includegraphics[width=.3\linewidth]{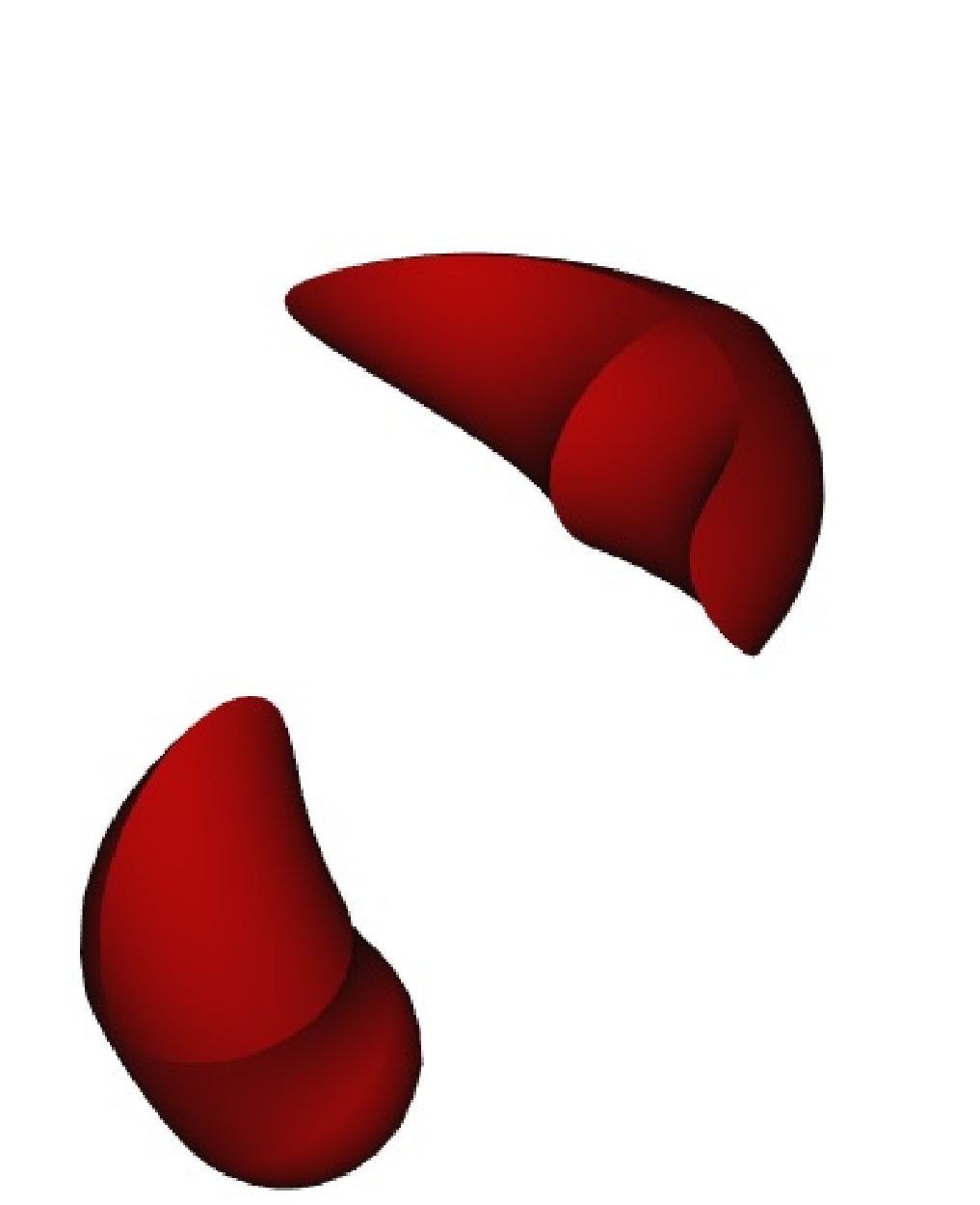}
\includegraphics[width=.3\linewidth]{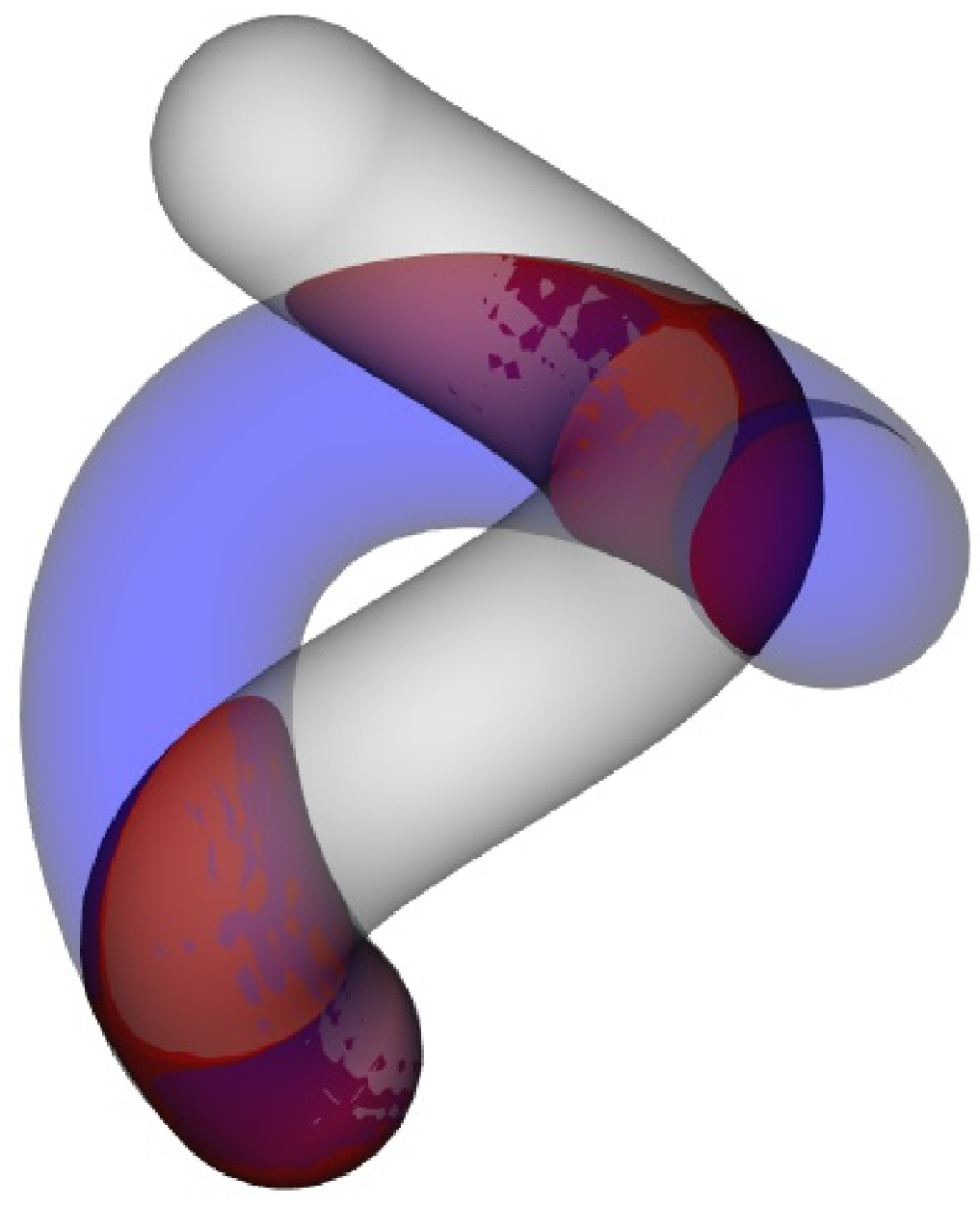}
\caption{
	Overlap (red) of the original helix (grey) and the transformed helix (blue) which is mirrored and rotated around the $y$-axis with $\Theta=125^\circ$ yielding the maximal overlap with $\chiGE=0.57$.
}
\label{fig:overlap}
\end{figure}

Similar to the approach of finding mirror planes in the $T$-matrix, mirror planes in a geometric object may be found by analyzing the overlap of the original object $O_r$
and its mirror image $O_l$. The mirror image $O_l$ is rotated by $(\Theta,\Phi)$ and translated from its centroid by the distance $d$ in order to maximize the overlap with the
original object $O_r$. In Fig.~\ref{fig:overlap}, we depict the original helix in grey and its rotated mirror image in blue. The overlap $O_l \cap O_r$ is shown in red.

For the geometric chirality coefficient $\chiGE$, the volume of the overlap $V(O_r \cap O_l)$ is maximized and given in units of the original volume $V(O_r)$
[\eqref{eq:chige} and Ref.~\cite{gilat1994}]. Since the mirror image of an achiral object may be brought to complete overlap, the coefficient vanishes in the achiral case: $\chiGE = 0$.
By contrast, if the mirror image does not overlap at all, the coefficient equals unity. Note, however, that there exists an overlap for any (including chiral) object.
Accordingly, $\chiGE=1$ is only a theoretical value.

In Fig.~\ref{fig:angleT}, the coefficient $\chiGE(\Theta,\Phi,0)$, which depends on the rotation angles, is shown. The relevant scalar, is the minimum of this coefficient as
it occurs for the maximal overlap. Including translations in $\chiGE(\Theta,\Phi,d)$, it is always possible to obtain $\chiGE(\Theta,\Phi,d) = 1$ by translating the mirror images
out of their respective bounding boxes. Only the analysis of all possible rotations and translations (i.e.~the minimization procedure)
yields a relevant coefficient which vanishes if the object possesses a mirror plane.

\subsection{Bayesian Optimization}

The global minimization of $\chiTT(\Theta,\Phi,d)$ in \eqref{eq:min} takes place in a three-dimensional space with parameters defined by the rotations around the angles
$(\Theta,\Phi)$ and the translation by the distance $d$.
Since the behaviour of the similarity of the original $T$-matrix $T_l$ and the matrix of the transformed mirror image $T_r$ is unknown, a global optimizer is
required to obtain the coefficient $\chiTT$ providing the highest symmetry in the $T$-matrix of the helix.

Bayesian Optimization (BO) is a procedure based on a stochastic model given by Gaussian processes (GPs) \cite{schneider2017}.
The optimization starts at a random point in the parameter space and predicts the objective function in the full space based on the previously obtained values.
This stochastic model is used to identify parameter values which yield a large expected improvement with respect to the currently known minimum.
Subsequently, the function value of the point with the highest expected improvement is determined and the predictive model is refined. 
Different stopping criteria such as the maximum number of function evaluations, the smallest probability of improvement or the smallest
expected improvement are possible.

\begin{figure}[htbp]
\centering
\includegraphics{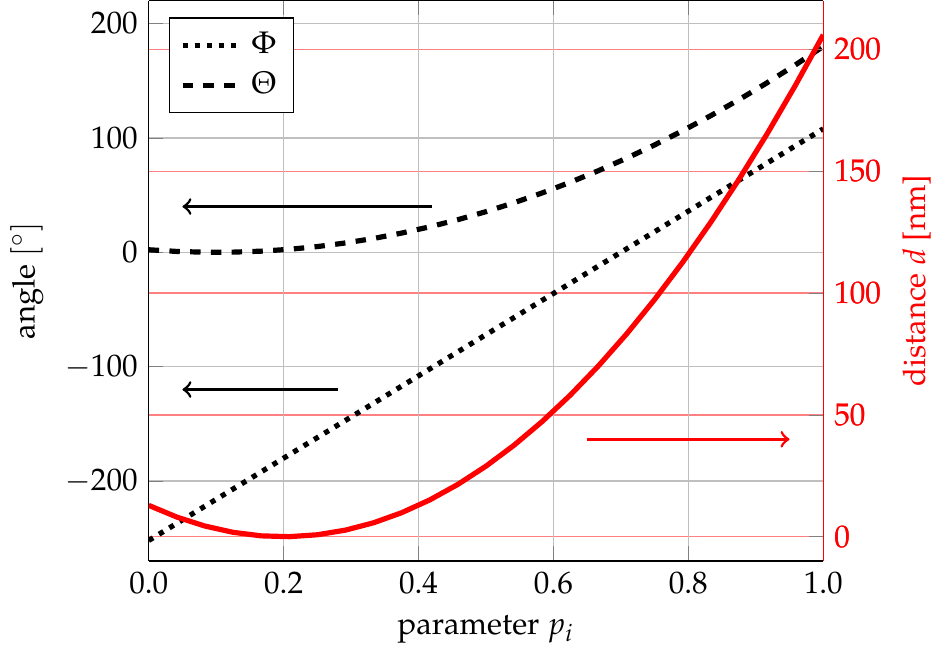}
\caption{
	Parametrization for optimization with $(p_1, p_2, p_3)$.
}
\label{fig:param}
\end{figure}

In the current study, we use a maximum number of 500 function evaluations, a minimal probability of improvement of $10^{-6}$ and a minimal expected improvement of $10^{-5}$.
The latter criterion is based on the numerical accuracy which is limited by the translation addition theorem for a finite $T$-matrix (cf.~Fig.~\ref{fig:transl}).
Further, we parametrize the space $(\Theta,\Phi,d)$ as follows in order to obtain parameters $(p_1, p_2, p_3) \in [0,1]^3$ and shift physical significant points such as
the centroid ($d=0$) to the inner part of the parameter intervals:
\begin{align}
	\Phi &= 360^\circ (p_1 - 0.7)
	, \\
	\Theta &= 180^\circ \left( \frac{p_2 - 0.1}{1 - 0.1} \right)^2
	, \\
	d &= 206\text{nm} \left( \frac{p_3 - 0.2}{1 - 0.2} \right)^2
	.
\end{align}
The quadratic functions involving $p_2$ and $p_3$ introduce ambiguities in the parameter space which are irrelevant for the physical values obtained from the inverse
functions
\begin{align}
	p_1 &= \frac{\Phi}{360^\circ} + 0.7
	, \\
	p_2 &= \sqrt{\frac{\Theta}{180^\circ}} (1.0 - 0.1) + 0.1
	, \\
	p_3 &= \sqrt{\frac{d}{206\text{nm}}} (1.0 - 0.2) + 0.2
	.
\end{align}
The relation between the variables $(\Theta,\Phi,d)$ and the parameters $(p_1, p_2, p_3)$ is depicted in Fig.~\ref{fig:param}.

\begin{figure}[htbp]
\centering
\includegraphics{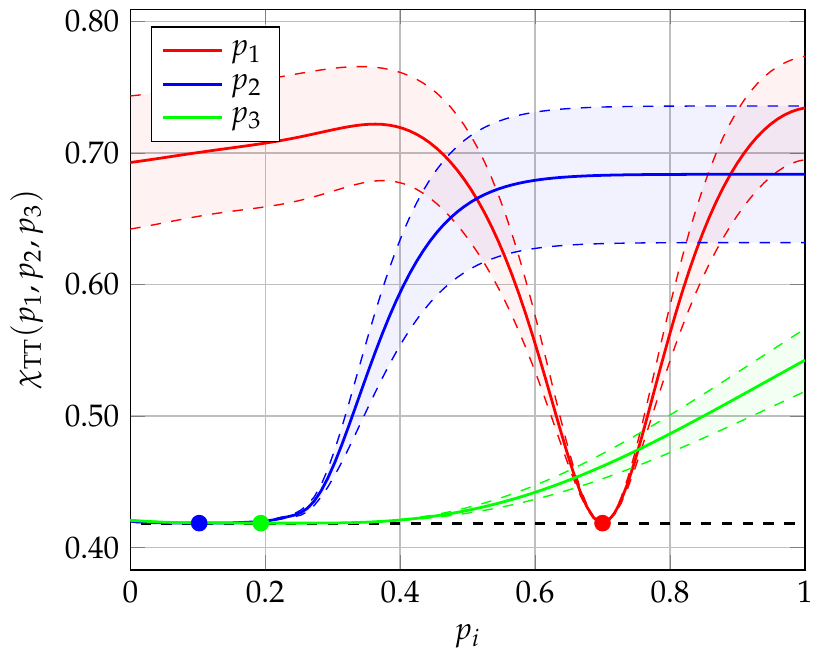}
\includegraphics{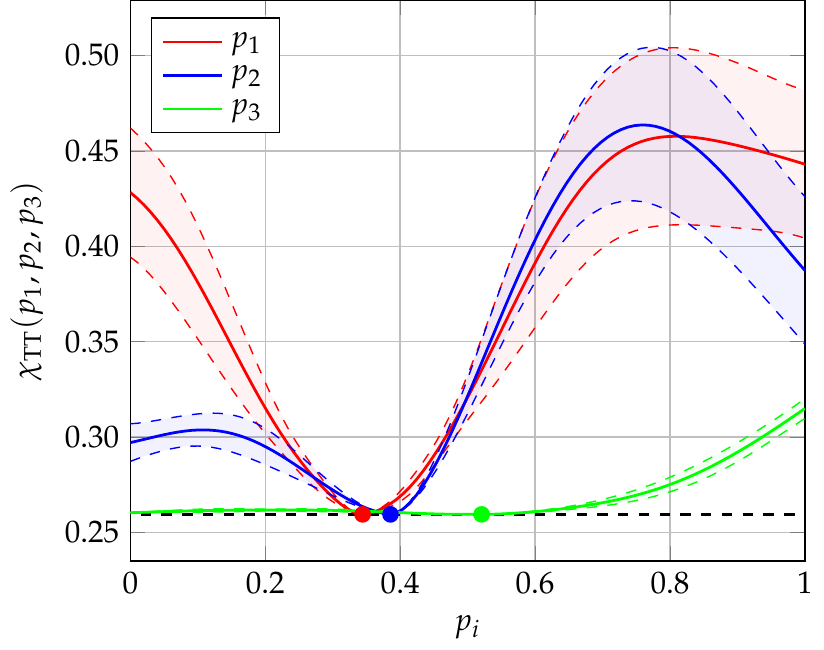}
\caption{
	Sensitivity of $\chiTT$ for $\lambda=823\text{nm}$ (top) and $\lambda=1,452\text{nm}$ (bottom).
	Minimal parameters (dots) and minimal $\chiTT$ (black dashed line) are shown.
	Solid lines are variations of $p_1$ (red), $p_2$ (blue) and $p_3$ (green) from the minimum.
	Shaded areas depict standard deviation derived from GPs.	
}
\label{fig:sens}
\end{figure}

The stochastic nature of the BO enables the predicition of objective values in the full parameter space as well as predicting the uncertainties, i.e.~standard deviations,
of these values. Accordingly, an interpolation of the parameter space is possible. For the two extrema of the CD, the prediction given by the BO is shown
in Fig.~\ref{fig:sens}. From the respective minimal values of $\chiTT(p_1^{\text{min}}, p_2^{\text{min}}, p_3^{\text{min}})$ cuts through the parameter space in all
three directions are shown. For each cut, one parameter (e.g.~$p_1$) is varied and the other parameters are kept constant (e.g.~$p_2=p_2^{\text{min}}$ and
$p_3 = p_3^{\text{min}}$).

As clearly visible, the minima are very flat with respect to parameter $p_3$, i.e.~translation from the centroid. For $\lambda=823\text{nm}$, the minimum is additionally
flat for $p_2$, i.e. rotations by $\Theta$.
In other words, small variations in the translation $d$ (and the rotation by $\Theta$ in the first case) do not change significantly $\chiTT$.
Accordingly, the similarity between the $T$-matrices of the mirror images do not change when $d$ (and $\Theta$) are varied.
That is why, the results of the optimization are subject to ambiguity caused by numerical fluctuations.

\begin{figure}[htbp]
\centering
\includegraphics{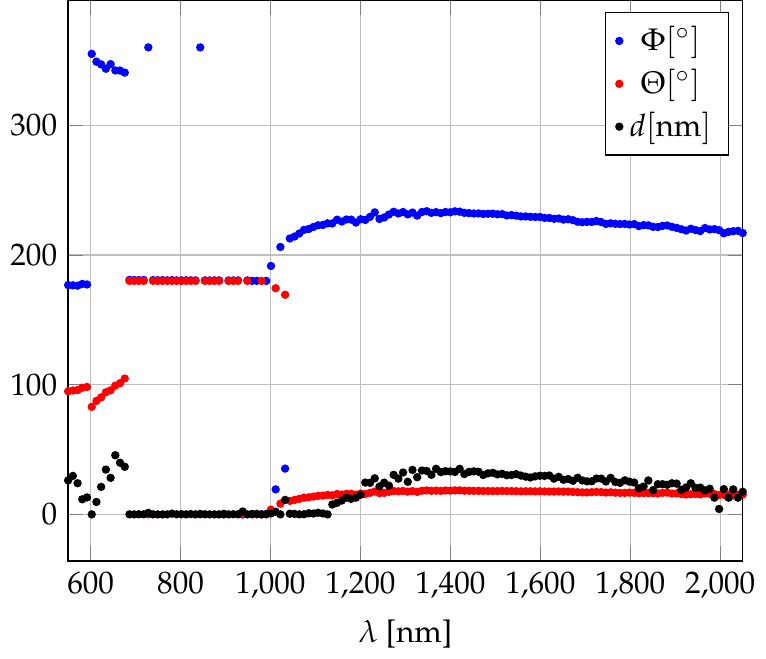}
\caption{
	Results of minimization.
}
\label{fig:mini}
\end{figure}

The results of the minimization are shown in Fig.~\ref{fig:mini}. These parameters correspond to the planes displayed in Fig.~\ref{fig:planes}.
As seen in the latter, the symmetry class 2 is divided into two subclasses: one class for $\lambda \in [600,680]\text{nm}$ and a second class for $\lambda \in [550,600]\text{nm}$.
These two classes differ only by a roation of $180^\circ$ about the $z$-axis. 

\end{document}